\documentclass[aps,pra,twocolumn,amsfonts,amssymb,amsmath,showpacs,showkeys,
floatfix,nofootinbib,groupedaddress,superscriptaddress,citesort]{revtex4}%
\usepackage{mathrsfs}
\usepackage{amsfonts}
\usepackage{amstext}
\usepackage{amsmath}
\usepackage{amssymb}
\usepackage{bm}
\usepackage{bbm}
\usepackage[dvips]{graphicx}
\def\qed{\leavevmode\unskip\penalty9999 \hbox{}\nobreak\hfill
     \quad\hbox{\leavevmode  \hbox to.77778em{%
              \hfil\vrule   \vbox to.675em%
               {\hrule width.6em\vfil\hrule}\vrule\hfil}}
     \par\vskip3pt}

\begin{document}

\preprint{APS/123-QED}
\title{Maximally coherent states}

\author{Zhaofang Bai}

\email{baizhaofang@xmu.edu.cn}\affiliation{School of Mathematical
Sciences, Xiamen University, Xiamen 361000, China}

\author{Shuanping Du}\thanks{Corresponding author}
\email{shuanpingdu@yahoo.com}\affiliation{School of Mathematical
Sciences, Xiamen University, Xiamen 361000,  China}



\begin{abstract}
The relative entropy measure quantifying coherence, a key property
of quantum system, is proposed recently. In this note, we firstly
investigate structural characterization of  maximally coherent
states with respect to the relative
entropy measure. 
It is shown that mixed maximally coherent states do not exist and
every pure maximally coherent state has the form
$U|\psi\rangle\langle \psi|U^\dag$,
$|\psi\rangle=\frac{1}{\sqrt{d}}\sum_{k=1}^{d}|k\rangle,$ $U$ is
diagonal unitary. Based on the characterization of pure maximally
coherent states, for a bipartite maximally coherent state with
$d_A=d_B$, we obtain that the super-additivity equality of
relative entropy measure holds if and only if the state is a
product state of its reduced states. From the viewpoint of
resource in quantum information, we find there exists a maximally
coherent state with maximal entanglement. Originated from the
behaviour of quantum correlation under the influence of quantum
operations, we further classify the incoherent operations which
send maximally coherent states to themselves.


\end{abstract}
\maketitle

\section{Introduction}
Being at the heart of interference phenomena, quantum coherence
plays a central role in physics as it enables applications that
are impossible within classical mechanics or ray optics. It
provides an important resource for quantum information processing,
for example, Deutsch's algorithm, Shor's algorithm, teleportation,
superdense coding and quantum cryptography \cite{Nielsen}.
Maximally coherent states are especially important for such
quantum information processing tasks.

Recently, it has attracted much attention to quantify the amount
of quantum coherence. In \cite{Bau}, the researchers establish a
quantitative theory of coherence as a resource following the
approach that has been established for entanglement in \cite{VP}.
They introduce a rigorous framework for quantification of
coherence by determining defining conditions for measures of
coherence and identifying classes of functionals that satisfy
these conditions. The relative entropy measure and $l_1$-norm
measure are proposed. Other potential candidates such as the
measures induced by the fidelity , $l_2$-norm and trace norm are
also discussed. It is shown that the coherence measure induced by
$l_2$-norm is not good. Since then, a lot of further
considerations about quantum coherence are stimulated
\cite{SXFL,XLF,MRS,MS,MC,RM,Gir,BDGX,BDX,Pire,Stre,BDQ}.

It has been shown that a good definition of coherence does not
only depend on the state of the system, but also depends on a
fixed basis for the quantum system \cite{Bau}. The particular
basis (of dimension $d$) we choose throughout this manuscript is
denoted by $\{|k\rangle\}_{k=1}^d$. In \cite{Bau}, Baumgratz etc.
identify the pure state
$|\psi\rangle:=\frac{1}{\sqrt{d}}\sum_{k=1}^{d}|k\rangle$ as a
maximally coherent state (MCS) with respect to any measure of
coherence
 because every state can be prepared from $|\psi\rangle$ by
a suitable incoherent operation. Two natural questions arise
immediately. Under a given coherence measurement, whether it is
the unique pure state whose coherence is maximal and whether there
exists a mixed maximally coherent state? Given a coherence measure
${\mathcal C}$, we call a state $\rho$ to be a maximally coherent
state (MCS) with respect to ${\mathcal C}$ if ${\mathcal C}(\rho)$
attends the maximal value of ${\mathcal C}$.

The relative entropy measure is able to not only quantify
coherence but also quantify superposition and frameness
\cite{Abe,Gour,Hor1,Vacc,Angel,Rodr}. In \cite{Fern}, the
regularized relative entropy measure of a resource can be used to
describe the optimal rate of converting (by asymptotically
resource non-generating operations) $n$ copies of a resource state
$\rho$ into $m$ copies of another resource state $\sigma$. On
considering the importance of the relative entropy measure, we are
aimed to characterize the structure of the maximally coherent
states under the relative entropy coherence measure. We obtain
that mixed maximally coherent states do not exist and each pure
maximally coherent state has the form $U|\psi\rangle$, where
$|\psi\rangle=\frac{1}{\sqrt{d}}\sum_{k=1}^{d}|k\rangle$, $U$ is
diagonal unitary. While it does not mean maximally coherent states
with respect to any coherence measure have the form
$U|\psi\rangle$. Indeed there exists a coherence measure such that
maximally coherent states with respect to this measure do not have
the form $U|\psi\rangle$ (see the example after Result 1).

Quantum correlation includes quantum entanglement and quantum
discord. Both entanglement and discord have a common necessary
condition---quantum coherence \cite{YZZ}. In \cite{XLF}, Z. Xi
etc. study the relative entropy coherence for a bipartite system
in a composite Hilbert space ${\mathcal H}^{AB}={\mathcal
H}^A\otimes {\mathcal H}^B$. They obtain an interesting property
for the relative entropy of coherence, that is, the
super-additivity,
 $$\mathcal C_{RE}(\rho)\geq
\mathcal C_{RE}(\rho_A)+\mathcal C_{RE}(\rho_B).$$ At the same
time, they leave an open question that whether the equality holds
if and only if $\rho=\rho_A\otimes \rho_B$. Using characterization
of MCS with respect to relative entropy coherence measure, we will
show that this question holds true if the two subsystems have the
same dimension and $\rho$ is a MCS. A counterexample is also given
to tell us that the answer is negative if the two subsystems have
different dimension. Furthermore, we obtain that there is a state
with maximal coherence and maximal entanglement.

Coherence,  as a kind of resource, enables applications that are
impossible within classical information. If an incoherent
operation  sends the MCSs to MCSs, we say it preserves MCSs.
 Naturally, does this kind of operation reduce the resource or is it a
without noise process? We will show  that an incoherent operation
preserves MCSs if and only if it has the form $U\cdot U^{\dag}$,
$U$ is a permutation of some diagonal unitary.

The structure of this paper is as follows.  Section II recalls the
axiomatic postulates for measures of coherence, the concepts of
the relative entropy measure and incoherent operations in
\cite{Bau}. In section III, we focus on the structural
characterization of maximally coherent states.  We apply this
characterization to bipartite system to answer the question on
super-additivity equality in section IV. The section V is devoted
to the incoherent operations preserving maximally coherent states.
The paper is ended with the conclusion in section VI.

\section{Preliminary}
Let ${\mathcal H}$ be a finite dimensional Hilbert space with
$d=\dim({\mathcal H})$. Fixing a basis $\{|k\rangle\}_{k=1}^d$, we
call all density operators (quantum states) that are diagonal in
this basis incoherent, and this set of quantum states will be
labelled by ${\mathcal I}$, all density operators $\rho\in
{\mathcal I}$ are of the form
$$\rho=\sum_{k=1}^d\lambda_k|k\rangle\langle k|.$$ Quantum
operations are specified by a finite set of Kraus operators
$\{K_n\}$ satisfying $\sum_n K_n^\dag K_n=I$, $I$ is the identity
operator on ${\mathcal H}$. From \cite{Bau}, quantum operations
are incoherent if they fulfil $K_n\rho K_n^\dag/Tr(K_n\rho
K_n^\dag)\in {\mathcal I}$ for all $\rho\in {\mathcal I}$ and for
all $n$. This definition guarantees that in an overall quantum
operation $\rho\mapsto \sum_nK_n\rho K_n^\dag$, even if one does
not have access to individual outcomes $n$, no observer would
conclude that coherence has been generated from an incoherent
state. Incoherent operations are of particular importance for the
decoherence mechanisms of single qubit \cite{Ava, Pre}. As a
special case, the unitary incoherent operation has the form
$\rho\mapsto U\rho U^\dag$, here $U$ is a permutation of a
diagonal unitary.

Based on Baumgratz et al.'s suggestion \cite{Bau}, any proper
measure of coherence ${\mathcal C}$ must satisfy the following
axiomatic postulates.

(i) The coherence vanishes on the set of incoherent states
(faithful criterion), ${\mathcal C}(\rho)=0$ for all
$\rho\in{\mathcal I}$;

(ii) Monotonicity under incoherent operation $\Phi$, ${\mathcal
C}(\Phi(\rho))\leq {\mathcal C}(\rho)$;

(iii) Non-increasing under mixing of quantum states (convexity),
$${\mathcal C}(\sum_np_n\rho_n)\leq \sum_np_n{\mathcal
C}(\rho_n)$$ for any ensemble $\{p_n,\rho_n\}$.

For any quantum state $\rho$ on the Hilbert space ${\mathcal H}$,
the measure of  relative entropy coherence   is defined as
$${\mathcal C}_{RE}(\rho):=\min_{\sigma\in{\mathcal
I}}S(\rho||\sigma),$$ where
$S(\rho||\sigma)=Tr(\rho\log_2\rho-\rho\log_2\sigma)$ is relative
entropy. In particular, there is  a closed form solution that
makes it easy to evaluate analytical expressions \cite{Bau}. For
Hilbert space ${\mathcal H}$ with the fixed basis
$\{|k\rangle\}_{k=1}^d$, we write
$\rho=\sum_{k,k'}p_{k,k'}|k\rangle\langle k'|$ and denote
$\rho_{diag}=\sum_kp_{kk}|k\rangle\langle k|$. By the properties
of relative entropy, it is easy to obtain $${\mathcal
C}_{RE}(\rho)=S(\rho_{diag})-S(\rho),$$ here $S(\cdot)$ is  von
Neumann entropy. Some basic properties of relative entropy
coherence have been given in \cite{Bau}.

Throughout the paper, if not specified, $\rho$ is a maximally
coherent state (MCS)  means that it is with respect to
 ${\mathcal C}_{RE}$. As we mention in introduction,
$|\psi\rangle:=\frac{1}{\sqrt{d}}\sum_{k=1}^{d}|k\rangle$ is a
maximally coherent state. That is, $\mathcal C_{RE}(|\psi\rangle
\langle \psi|)=\log_2 d$ is the maximal value of $\mathcal
C_{RE}$.  The structural characterization of MCS plays a key role
in section IV and V. An incoherent operation $\Phi$ preserves MCSs
means that $\Phi(\rho)$ is a MCS if $\rho$ is a MCS.

\section{Maximally coherent states on ${\mathcal H}$}

\textbf{ Result 1.} $\rho$ is a MCS if and only if
$\rho=U|\psi\rangle\langle \psi |U^\dag$, where
$|\psi\rangle=\frac{1}{\sqrt{d}}\sum_{k=1}^{d}|k\rangle$ and $U$
is a diagonal unitary.

\textbf{ Proof.} It is easy to see, for every diagonal unitary
element $U$, $\rho\mapsto U\rho U^{\dag}$ is an incoherent
operation. From the monotonicity under incoherent operations, it
follows that $U|\psi\rangle\langle \psi |U^{\dag}$ is a MCS.

For if part, we firstly prove that $\rho$ is pure. Note that the
maximal value of $\mathcal C_{RE}$ is  $\log_2 d$. For every pure
state ensemble $\rho=\sum_i p_i\rho _i$. If $\mathcal
C_{RE}(\rho)=\log_2 d$, then $$\log_2 d=\mathcal C_{RE}(\rho)
\leq\sum_{i}p_i\mathcal C_{RE}(\rho_i)\leq \log_2 d.$$ Thus
$\mathcal C_{RE}(\rho) =\sum_{i}p_i\mathcal C_{RE}(\rho_i)$ and
$\mathcal C_{RE}(\rho_i)= \log_2 d$. Let  $\mathcal
C_{RE}(\rho_i)=S(\rho_i\parallel\sigma_i)$ and
$\sigma=\sum_{i}p_i\sigma_i$. By the jointly convex of relative
entropy, $$\log_2 d\leq S(\rho||\sigma)\leq\sum_i p_iS(\rho_i
||\sigma_i)= \log_2 d.$$ This implies  $S(\rho||\sigma)=\sum_i
p_iS(\rho_i||\sigma_i)$. From \cite[Theorem 10]{Ana}, it follows
that $\rho_i=\rho_j$ and so $\rho$ is a pure state.

Now, we write $\rho=|\phi\rangle\langle \phi|$ and $|\phi\rangle
=\sum_{k=1}^{d}\alpha_{k}|k\rangle$. By the property of relative
entropy, $$\mathcal C_{RE}(\rho)=S(\rho_{diag})=-\sum_{
k=1}^{d}|\alpha_k|^2\log_2(|\alpha_k|^2).$$  A direct computation
shows that  $\mathcal C_{RE}(\rho)=\log_2(d)$ implies that
$|\alpha_k|^2=1/d$. One can write
$\alpha_k=\frac{1}{\sqrt{d}}e^{i\theta_k}$, then $|\phi\rangle
=\sum_{k=1}^{d}\frac{1}{\sqrt{d}}e^{i\theta_k}|k\rangle$. Let
$U=\text {diag}(e^{i\theta_1}, e^{i\theta_2}, ...,
e^{i\theta_d})$, so $|\phi\rangle=U|\psi\rangle$. $\square$

In \cite{Bau}, it is mentioned that if $\mathcal D$ is distance
measure satisfying contracting under CPTP maps and jointly convex,
(i.e., satisfying  $\mathcal D(\rho,\sigma)\geq \mathcal
D(\Phi_{CPTP}(\rho),\Phi_{CPTP}(\sigma))$ and $\mathcal D(\sum_n
p_n\rho_n, \sum_n p_n \sigma_n)\leq \sum_n p_n\mathcal
D(\rho_n,\sigma_n))$, then one may define a coherence measure by
$$\mathcal{C}_{\mathcal D}(\rho)=\min_{\sigma\in {\mathcal
I}}\mathcal D(\rho,\sigma).$$ From the proof of Result 1, it is
easy to see that if  $\mathcal D$ possesses the property that the
equality of jointly convex holds true implies $\rho_n=\rho_{m}$,
then the MCSs with respect to the coherence measure induced by
$\mathcal D$ are pure. It is known that $l_1$-norm \cite{Bau} and
quantum skew divergence \cite{Aud} are with such property.

Here we remark that Result 1 does not hold true for any coherence
measure. The following is a counter example.

{\bf Example.} Let $d=4$ and $\Omega=\{{\bf x}=( x_1,
x_2,x_3,x_4)^t\mid \sum_{i=1}^4 x_i=1 \text{ and } x_i\geq 0\} $,
here $( x_1, x_2, x_3,x_4)^t$ denotes the transpose of row vector
$( x_1, x_2,x_3,x_4)$. Assume
$$f({\bf x})=\left\{\begin{array}{ll}
                                             -\sum_{i=1}^4x_i\log_2
                                             x_i,&x_4^\downarrow=0\\
                                             \log_2 3,&
                                             x_4^\downarrow\neq 0\end{array}\right.,$$
here $x_4^\downarrow$ is the least element in $(x_1,x_2,
x_3,x_4)^t$. By \cite[Theorem 1]{BDQ}, it is easy to check that
the nonnegative function $f$ can derive a coherence measure $C_f$.
It is clear that both
$|\psi\rangle=\sum_{k=1}^4\sqrt{x_k}|k\rangle,x_4^\downarrow\neq
0$ and $|\phi\rangle=\sum_{k=1}^3\sqrt{\frac{1}{3}}|k\rangle$ are
maximally coherent  under $C_f$.

\section{Maximally coherent states  on $H_A\otimes H_B$}

Consider a bipartite system in a composite Hilbert space
${\mathcal H}^{AB} = {\mathcal H}^A\otimes {\mathcal H}^B$ of $d =
d_A \times d_B$ dimension, here $d_A = \dim({\mathcal H}^A)$ and
$d_B =\dim({\mathcal H}^B)$. Let $\{|k\rangle^A\}_{k=1}^{d_A}$ and
$\{|j\rangle ^B\}_{j=1}^{d_B}$ be the orthogonal basis for the
Hilbert space ${\mathcal H}^A$ and ${\mathcal H}^B$, respectively.
Given a quantum state $\rho_{AB}$ which could be shared between
two parties, Alice and Bob, and let $\rho_A$ and $\rho_B$ be the
reduced density operator for each party.

In \cite{XLF}, Xi etc. show the supper-additivity of the relative
entropy coherence:
\begin{equation}\label{1}\mathcal C_{RE}(\rho_{AB})\geq
\mathcal C_{RE}(\rho_A)+\mathcal C_{RE}(\rho_B).\end{equation}
They leave an question that whether the equality holds if and only
if $\rho=\rho_A\otimes \rho_B$. 
In the following, we will show that the answer is affirmative if
$d_A=d_B$ and $\rho_{AB}$ is a MSC. If $d_A\neq d_B$, then the
answer is negative. This implies that, in the case of $d_A=d_B$,
there is a correlation between the two subsystems, this leads to
the increase of the coherence on the bipartite system.

\textbf{Result 2.} If $d_A=d_B$ and $\rho_{AB}$ is a MCS, then the
equality in (\ref{1}) holds if and only if
$\rho_{AB}=\rho_A\otimes \rho_B$.

\textbf{ Proof.} Let $\{|i\rangle^A\}_{i=1}^{d_A}$ and
$\{|j\rangle ^B\}_{j=1}^{d_B}$ be the orthogonal basis for the
Hilbert space $H_A$ and $H_B$, respectively. Let
$\rho_{AB}=|\phi\rangle\langle \phi|$ with $|\phi\rangle =\frac 1
{\sqrt{d} }\sum_{i,j=1}^{d_A,d_B} e^{i\theta_{ij}}|i^Aj^B\rangle$.
Then \begin{equation}\label{7}\rho=\frac 1 {d} \sum_{i,j,s,t}
e^{i(\theta_{ij}-\theta_{st})}) |i^A\rangle \langle s^A|\otimes
|j^B\rangle \langle t^B|, \end{equation}
$$\rho_A=\frac 1 {d}
\sum_{i,s}(\sum_j e^{i(\theta_{ij}-\theta_{sj})}) |i^A\rangle
\langle s^A|$$ and
$$\rho_B=\frac 1 {d} \sum_{j,t}(\sum_i
e^{i(\theta_{ij}-\theta_{it})}) |j^B\rangle \langle t^B|.$$ Note
that $\mathcal C_{RE}(\rho_{AB})= \mathcal C_{RE}(\rho_A)+\mathcal
C_{RE}(\rho_B)\Leftrightarrow\rho_A,\rho_B$ are MCSs
$\Leftrightarrow|\sum_i e^{i(\theta_{ij}-\theta_{it})}|=d_A$ and
$|\sum_j e^{i(\theta_{ij}-\theta_{sj})}|=d_B$. The latter
equivalence follows from Result 1. By a direct computation, we
have
\begin{equation}\label{4}\theta_{ij}-\theta_{it}=\theta_{i'j}-\theta_{i't}\text{
and }
\theta_{ij}-\theta_{sj}=\theta_{ij'}-\theta_{sj'}.\end{equation}
On the other hand, \begin{equation}\label{8}\rho_A\otimes \rho_B =
\frac 1 {d} \sum_{i,j,s,t} \alpha_{ijst}
 |i^A\rangle \langle s^A|\otimes
|j^B\rangle \langle t^B|,\end{equation}  here
$\alpha_{ijst}=\frac{1}{d}(\sum_j
e^{i(\theta_{ij}-\theta_{sj})})(\sum_i
e^{i(\theta_{ij}-\theta_{it})}).$ From Equations
(\ref{7}),(\ref{4}) and (\ref{8}), we finish the proof. $\square$

What will happen if $d_A\neq d_B$? The following counterexample
shows the answer is negative in this case.

Assume $d_A=2$ and $d_B=3$. Let $$|\phi\rangle=\frac 1
{\sqrt{6}}(|1\rangle+e^{i\theta}|2\rangle+e^{2
i\theta}|3\rangle+e^{3i\theta}|4\rangle+e^{4i\theta}|5\rangle+e^{5i\theta}|6\rangle),$$
$\theta\in(0,2\pi)$. Clearly, $\rho=|\phi\rangle\langle \phi|$ is
a MCS. By an elementary computation,
$$\rho=\frac 1 6\left(\begin{array}{cccccc}
1 & e^{-i\theta} & e^{-2i\theta}
&e^{-3i\theta} &e^{-4i\theta} &e^{-5i\theta}\\
e^{i\theta} & 1 & e^{-i\theta}
&e^{-2i\theta} & e^{-3i\theta} & e^{-4i\theta}\\
e^{2i\theta} & e^{i\theta} & 1 & e^{-i\theta} &e^{-2i\theta}
&e^{-3i\theta}\\
e^{3i\theta} & e^{2i\theta} & e^{i\theta} &1 &e^{-i\theta}
&e^{-2i\theta}\\
e^{4i\theta} & e^{3i\theta} & e^{2i\theta} &e^{i\theta} &1
&e^{-i\theta}\\
 e^{5i\theta} & e^{4i\theta} &e^{3i\theta} &e^{2i\theta}
&e^{i\theta}&1\\
\end{array}\right).$$
$$\rho_A=\frac 1 2\left(\begin{array}{cc}
1& e^{-3i\theta}\\
e^{3i\theta} &1 \\
\end{array}\right),$$
$$\rho_B=\frac 1 3\left(\begin{array}{ccc}
1& e^{-i\theta} & e^{-2i\theta}\\
e^{i\theta}& 1 & e^{-i\theta}\\
e^{2i\theta}& e^{i\theta} & 1
\end{array}\right).$$
It is evident that both $\rho_A$ and $\rho_B$ are MCSs and
$\mathcal C_{RE}(\rho_{AB})=\mathcal C_{RE}(\rho_A)+\mathcal
C_{RE}(\rho_B)$, however $\rho\neq \rho_A\otimes \rho_B$.

It is wellknown that both coherence and entanglement are
considered as resource in quantum information. Whether is there a
state which is not only maximally coherent but also maximally
entangle? We will discuss this important question at the end of
this section.

\textbf{Result 3.} There is a MCS $\rho$ which is maximal
entanglement.

\textbf{ Proof.} Let $\rho=|\phi\rangle\langle \phi|$ with
$$|\phi\rangle =\frac 1 {\sqrt{d} }\sum_{i,j=1}^{d_A,d_B}
e^{i\theta_{ij}}|i^Aj^B\rangle.$$ Then $\rho_A=\frac 1 d
\sum_{i,s}(\sum_j e^{i(\theta_{ij}-\theta_{sj})}) |i^A\rangle
\langle s^A|$. Recall that $\rho$ is  maximally entangled if and
only if $\rho_A=\frac I {d_A}$. Therefore
\begin{equation}\label{*} \sum_j e^{i(\theta_{ij}-\theta_{sj})}=0
\text{ for every pair }i\neq s \end{equation} implies that $\rho$
is a maximally entangle state. Note that The equation (\ref{*})
has a solution. In order to understand the solution, we list an
example in the case of $d_A=d_B=3$.
$\theta_{11}=\theta_{12}=\theta_{13}=0$, $\theta_{21}=0$,
$\theta_{22}=-\frac{2\pi}3$, $\theta_{23}=-\frac {4\pi} 3$,
$\theta_{31}=0$, $\theta_{32}=-\frac{4\pi}3$, and
$\theta_{33}=-\frac{2\pi}3$. $\square$

\section{Incoherent operations preserving MCS}

It is an interesting area to study the behavior of quantum
correlation under the influence of quantum operations
~\cite{Streltsov,Filippov,Zyczkowski2,Rao,Cui,Shabani,Altinatas,Mazzola,Ciccarello,
Hu,Guo3,Guo1,Guo2,Hassan,GBD,BD}. For example, local operations
that cannot create QD is investigated in \cite{Streltsov,Hu,Guo1},
local operations that preserve the state with vanished MIN is
characterized in \cite{Guo3} and local operations that preserve
the maximally entangled states is explored in \cite{Guo2}. The
goal of this chapter is to discuss when an incoherent operation
preserves MCSs.

Here is our main result in this section.

 \textbf{Result 4.} An incoherent operation $\Phi$
preserves MCSs if and only if $\Phi(\rho)=U\rho U^{\dag}$ for
every quantum state $\rho$, here $U$ is a permutation of a
diagonal unitary.

From Result 4, every incoherent operation preserving  MCSs does
not reduce the resource and is noiseless. Although this result is
not surprising, the proof is not trivial. Let $\Phi$ be specified
by a set of Kraus operators $\{K_n\}$, the main step of our proof
is to show that each $K_n=a_n\Pi_n$ after some reduction, $a_n$ is
a complex number with $\sum_n |a_n|^2=1$ and $\Pi_n$ is a
permutation of $I$. The reduction process is not trivial because
we need to prove $\Phi$ is unital which is based on an interesting
property that identity operator can be described as a sum of $d$
MCSs.

\textbf{Proof.} The if part can be obtained directly from the
Result 1.

Now we check the only if part. We firstly claim that $I$ can be
written as $\sum_{k=1}^d |\phi_k\rangle\langle \phi_k|$ with all
of $|\phi_k\rangle$ are MCSs. Choose
$|\phi_j\rangle=\frac{1}{\sqrt{d}}\sum_{k=1}^d
e^{i\alpha_{j,k}}|k\rangle$, all of $\alpha_{j,k}$ are real
numbers. Denote $M=\sum_{j=1}^d|\phi_j\rangle\langle \phi_j|$,
then $M$ has the matrix form \begin{widetext}
$$\left(\begin{array}{cccc}
  1 & \frac 1 d
  \sum_{j=1}^d e^{i(\alpha_{j,1}-\alpha_{j,2})}& \cdots
  &   \frac 1 d \sum_{j=1}^d e^{i(\alpha_{j,1}-\alpha_{j,d})}\\
\frac 1 d \sum_{j=1}^d e^{i(\alpha_{j,2}-\alpha_{j,1})} & 1 &
\cdots& \frac 1 d \sum_{j=1}^d e^{i(\alpha_{j,2}-\alpha_{j,d})}\\
\cdots & \cdots &
\cdots &\cdots\\
\frac 1 d  \sum_{j=1}^d e^{i(\alpha_{j,d}-\alpha_{j,1})} & \frac 1
d \sum_{j=1}^d e^{i(\alpha_{j,d}-\alpha_{j,2})}& \cdots &
1\end{array}\right).$$
\end{widetext}
If $\alpha_{j,k}$ satisfy
$$\alpha_{j+1,k}-\alpha_{j+1,l}=\alpha_{j,k}-\alpha_{j,l}+\frac {2(k-l)}{d} \pi,$$
then $\sum_{j=1}^d e^{i(\alpha_{j,k}-\alpha_{j,l})}=0$
($j,k,l=1,\ldots,d$,  $k\neq l$). So $M=I$. There exist solutions
of these equations, for example $\alpha_{j,k}=\frac 2 d
(k-1)(j-1)\pi$.

In the following, we show that  $\Phi$ preserving MCS is unital,
that is $\Phi(I)=I$. Note that $\Phi$ is incoherent, we have
$\Phi(I)$ is diagonal. From the Result 1 in section III,
$\Phi(|\phi_k\rangle\langle \phi_k|)=U_k |\psi\rangle\langle
\psi|U_k^{\dag}$, $U_k$ is diagonal unitary. Then
$(\Phi(|\phi_k\rangle\langle \phi_k|))_{diag}=\frac{I}{d}$, here
$(\Phi(|\phi_k\rangle\langle \phi_k|))_{diag}$ denotes the state
obtained from $\Phi(|\phi_k\rangle\langle \phi_k|)$ by deleting
all off-diagonal elements.  This implies that
$$\Phi(I)=\Phi(I)_{diag}=\sum_{k=1}^d (\Phi(|\phi_k\rangle\langle
\phi_k|))_{diag}=I.$$

Let $K_n$ be the Kraus operators of $\Phi$, we obtain $\sum_n
K_nK_n^{\dag}=\sum_n K_n^{\dag}K_n=I$. From $\Phi$ is incoherent,
we also have that every column of $K_n$ is with  at most 1 nonzero
entry. From Result 1, for every  diagonal unitary $U$, there is a
diagonal unitary $V_U$ depending on $U$ such that
$\Phi(U|\psi\rangle\langle \psi|U^{\dag})=V_U|\psi\rangle\langle
\psi|V_U^{\dag}$.  That is $|\psi\rangle\langle \psi|$ is a fixed
point of $V_U^{\dag}\Phi(U\cdot U^{\dag})V_U$. This implies  $$V_U
^{\dag}K_n U|\psi\rangle\langle \psi|=|\psi\rangle\langle \psi|V_U
^{\dag}K_n U.$$ So $V_U ^{\dag}K_n U
|\psi\rangle=\lambda_{n,U}|\psi\rangle$ for some scalar $\lambda
_{n,U}$ depending on $U$ and $n$. We assert that
$\lambda_{n,I}\neq0$. Otherwise, $K_n$ is singular and so there is
a row of $K_n$ in which all entries are zero. Note that
$|\psi\rangle =\frac{1}{\sqrt{d}}\sum_{k=1}^{d}|k\rangle$,
therefore all $\lambda_{n,U}$ equal zero and so $K_nU|\psi|=0$.
Since $I$ can be written as a sum of MCSs, we have $K_n=0$. From
$\lambda_{n,I}\neq0$, there exists a nonzero element of each row
of $K_n$. Combining this and each column of $K_n$ is with at most
one nonzero element, we get that there is one and only one nonzero
entry in every row and column of $K_n$. Note that $V_I^\dag\Phi
V_I$ possesses the same properties as $\Phi$, without loss of
generality, we may assume $\Phi(|\psi\rangle\langle
\psi|)=|\psi\rangle\langle \psi|.$ So $K_{n}
|\psi\rangle=\lambda_{n,I}|\psi\rangle$. This implies the entries
of $K_n$ are equal. Therefore $K_n=a_n \Pi_n$, $a_n$ is a complex
number with $\sum_n |a_n|^2=1$ and $\Pi_n$ is a permutation of
$I$.

From Result 1, for arbitrary $d$ real numbers $\theta_1,\cdots
,\theta_d$, $|\phi\rangle=\sum_k \frac 1{\sqrt{d}}
e^{i\theta_k}|k\rangle$ is a MCS. By a direct computation, $K_n
|\phi\rangle=\frac{a_n}{\sqrt{d}}\sum_k
e^{i\alpha_{kn}}|k\rangle$,
$(\alpha_{1n},\cdots,\alpha_{dn})=\Pi_n(\theta_1,\cdots,\theta_d)$.
Furthermore, $K_n|\phi\rangle\langle \phi|K_n^{\dag}$ is the
matrix
$$\frac{|a_n|^2}{d}\left(\begin{array}{cccc}
  1 & e^{i(\alpha_{1n}-\alpha_{2n})} & \cdots & e^{i(\alpha_{1n}-\alpha_{dn})}\\
e^{i(\alpha_{2n}-\alpha_{1n})} & 1 & \cdots & e^{i(\alpha_{2n}-\alpha_{dn})}\\
\cdots & \cdots&\cdots &\cdots\\
e^{i(\alpha_{dn}-\alpha_{1n})} &e^{i(\alpha_{dn}-\alpha_{2n})} &
\cdots& 1\end{array}\right).$$ And $\Phi(|\phi\rangle\langle
\phi|)$ equals
\begin{widetext} $$\frac{1}{d}\left(\begin{array}{cccc}
  \sum_n |a_n|^2 & \sum_n |a_n|^2e^{i(\alpha_{1n}-\alpha_{2n})} & \cdots & \sum_n |a_n|^2e^{i(\alpha_{1n}-\alpha_{dn})}\\
\sum_n |a_n|^2e^{i(\alpha_{2n}-\alpha_{1n})} & \sum_n |a_n|^2 & \cdots & \sum_n |a_n|^2e^{i(\alpha_{2n}-\alpha_{dn})}\\
\cdots & \cdots&\cdots &\cdots\\
\sum_n |a_n|^2e^{i(\alpha_{dn}-\alpha_{1n})} &\sum_n
|a_n|^2e^{i(\alpha_{dn}-\alpha_{2n})} & \cdots& \sum_n
|a_n|^2\end{array}\right).$$ \end{widetext}

By our assumption, it is a MCS. So $$|\sum_n
|a_n|^2e^{i(\alpha_{jn}-\alpha_{kn})}|=1$$ for $j,k=1,2,\cdots,
d$. The arbitrariness of $ \alpha_{jn}$ and $\alpha_{kn}$ implies
$n=1$. Therefore $\Phi$ has the desired form. $\square$

\section{conclusion}

In this paper, we firstly investigate the maximally coherent
states with respect to the relative entropy measure of coherence.
We find that there does not exist a mixed maximally coherent state
and each pure maximally coherent states have the form
$U|\psi\rangle$, where $U$ is a
 diagonal unitary and
$|\psi\rangle:=\frac{1}{\sqrt{d}}\sum_{k=1}^{d}|k\rangle$.
Applying this structural characterization of maximally coherent
states to bipartite system, we
 answer the question left in \cite{XLF} whether $\mathcal
C_{RE}(\rho_{AB})= \mathcal C_{RE}(\rho_A)+\mathcal
C_{RE}(\rho_B)$ if and only $\rho=\rho_A\otimes \rho_B$. It is
shown that the answer is affirmative if $d_A=d_B$ and $\rho_{AB}$
is a MSC. If $d_A\neq d_B$, then the answer is negative. From the
viewpoint of resource of quantum information, we show that there
exists a state which is not only maximally coherent but also
maximally entangled. By using the form of pure maximally coherent
states, we obtain the structural characterization of incoherent
operations sending maximally coherent states into maximally
coherent states. That is, an incoherent operation $\Phi$ preserves
MCSs if and only if $\Phi(\rho)=U\rho U^{\dag}$ for every quantum
state $\rho$, here $U$ is a permutation of a diagonal unitary.

\section{Acknowledgement}
This work was completed while the authors were visiting the
Department of Mathematics and Statistics of the University of
Guelph and IQC of the University of Waterloo during the academic
year 2014-2015 under the support of China Scholarship Council. We
thank Professor David W. Kribs and Professor Bei Zeng for their
hospitality. This work is partially supported by the Natural
Science Foundation of China (No. 11001230), and the Natural
Science Foundation of Fujian (2013J01022, 2014J01024).


\begin{thebibliography}{99}
\bibitem{Nielsen}M. A. Nielsen and I. L. Chuang, Quantum Computation and Quantum
information (Cambridge University Press, Cambridge, 2000).

\bibitem{Bau}T. Baumgratz, M. Cramer, and M.B. Plenio, Phys. Rev. Lett. 113, 140401
(2014).

\bibitem{VP}V. Vedral and M.B. Plenio, Phys. Rev. A 57, 1619 (1998).

\bibitem{XLF}Z. Xi, Y. Li and H. Fan, arXiv:1408.3194v1.

\bibitem{RM} \'{A}. Rivas and M. M\"{u}ller arXiv:1409.1770.

\bibitem{MRS}I. Marvian, Robert W. Spekkens, Phys. Rev. A \textbf{90},
062110 (2014).

\bibitem{MS} Iman Marvian and RobertW. Spekkens, Nat. Commun, \textbf{5}, 3821
(2014).

\bibitem{MC} A. Monras, A.Ch\c{e}ci\'{n}ska and A. Ekert, New J.
Phys.\textbf{16} 063041 (2014).



\bibitem{Gir} D. Girolami, Phys. Rev. Lett. \textbf{113}, 170401 (2014).

\bibitem{BDGX} S. P.  Du, Z. F. Bai and Y. Guo, Phys. Rev. A
\textbf{91}, 052120 (2015).

\bibitem{BDX} S. P.  Du, Z. F. Bai,  Annals of Phys. \textbf{359},
136 (2015).

\bibitem{SXFL} L. H. Shao, Z. J. Xi, H. Fan and Y. M. Li, Phys. Rev. A \textbf{91}, 042120
(2015).

\bibitem{Pire} D. P. Pires, L. C. Celeri, D. O. Soares-Pinto,
arXiv:1501.05271.

\bibitem{Stre} A. Streltsov, U. Singh, H. S. Dhar, M. N. Bera, and
G. Adesso, arXiv:1502.05876.

\bibitem{BDQ} S. P.  Du, Z. F. Bai and X. F. Qi,
arXiv:1504.02862v1.

\bibitem{Hor1} M. Horodecki, P. Horodecki, R. Horodecki, J. Oppenheim, A. Sen,
U. Sen, and B. Synak-Radtke, Phys. Rev. A \textbf{71}, 062307
(2005).


\bibitem {Abe} J. $\AA$berg, arXiv:quant-ph/0612146.

\bibitem{Vacc}  J. A. Vaccaro, F. Anselmi, H. M. Wiseman, and K. Jacobs, Phys.
Rev. A \textbf{77}, 032114 (2008).


\bibitem{Gour} G. Gour, I. Marvian, and R. W. Spekkens, Phys. Rev. A \textbf{80}, 012307
(2009).

\bibitem{Rodr} C. A. Rodr\'{i}guez-Rosario, T. Frauenheim, and A. Aspuru-Guzik,
arXiv:1308.1245.


\bibitem{Angel} R. M. Angelo and A. D. Ribeiro, Found. Phys. DOI:10.1007/s10701-015-9913-6.




\bibitem{Fern} Fernando G. S. L. Brand\~{a}o and  Gilad Gour, arXiv:1502.03139v1.

\bibitem{YZZ}  C. S. Yu, Y. Zhang and H. Q. Zhao, Quantum Information Processing. \textbf{13}, 1437
(2014).


\bibitem{Ava} M. Avalle, A. Serafini, Phys. Rev. Lett. \textbf{112}, 170403(2014).

\bibitem{Pre} J. Preskill,  
(Lecture Notes for Physics 229, California Institute of
Technology, 1998).

\bibitem{Ana} A. Jen\v{c}ov\'{a} and M. B. Ruskai, 
 Rev. Math. Phys., \textbf{22}, 1099 (2010).

\bibitem{Aud} K. M. R. Audenaert, 
J. Math. Phys., \textbf{55}, 112202 (2014).









\bibitem{Zyczkowski2}
 K. \.{Z}yczkowski, P. Horodecki, M. Horodecki, and R. Horodecki:
{\em Phys. Rev. A} {\textbf{65}}, 012101 (2001).

\bibitem{Shabani}
A. Shabani and D. A. Lidar: {\em Phys. Rev. Lett.} {\textbf{102}},
100402 (2009).

\bibitem{Cui} W. Cui, Z. Xi, and Y. Pan:
{\em J. Phys. A: Math. Theor.} {\textbf{42}}, 155303 (2009).

\bibitem{Altinatas}
F. Altintas and R. Eryigit: {\em J. Phys. A: Math. Theor.}
{\textbf{43}}, 415306 (2010).

\bibitem{Mazzola} L. Mazzola, J. Piilo, and S. Maniscalco:
{\em Phys. Rev. Lett.} {\textbf{104}}, 200401 (2010).


\bibitem{Rao}
B. R. Rao, R. Srikanth, C. M. Chandrashekar, and S. Banerjee: {\em
Phys. Rev. A} {\textbf{83}}, 064302 (2011).

\bibitem{Streltsov} A. Streltsov, H. Kampermann, and D. Bru{\ss}:
{\em Phys. Rev. lett.} \textbf{107}, 170502 (2011).

\bibitem{Ciccarello} F. Ciccarello and V. Giovannetti:
{\em Phys. Rev. A} {\textbf{85}}, 010102 (2012).

\bibitem{Filippov} S. N. Filippov, T. Ryb\'{a}r, and M. Ziman:
{\em Phys. Rev. A} {\textbf{85}}, 012303 (2012).

\bibitem{Hu} X. Hu, H. Fan, D. Zhou, and W. Liu:
{\em Phys. Rev. A} {\textbf{85}}, 032102 (2012).

\bibitem{Guo3}Y. Guo and J. Hou:
 {\em J. Phys. A: Math. Theor.} \textbf{46}, 325301 (2013).

\bibitem{Guo1}
Y. Guo and J. Hou:
{\em J. Phys. A: Math. Theor.} \textbf{46}, 155301 (2013).

\bibitem{Guo2}Y. Guo, Z. Bai, and S. Du:
{\em Int. J. Theor. Phys.}  \textbf{52}, 3820--3829 (2013).

\bibitem{Hassan} A. S. M. Hassan and P. S. Joag:
{\em Eur. Phys. Lett.} \textbf{103}, 10004 (2013).


\bibitem{GBD} Y. Guo, Z. Bai and S. Du: 
{\em Rep. Math. Phy.} \textbf{74}, 277 (2014).

\bibitem{BD} Z. Bai and S. Du: 
{\em J. Phys. A: Math. Theor.} \textbf{47}, 175302
(2014).\vspace{1in}

\end{thebibliography}
\end{document}